# Embedding parameters in *ab initio* theory to develop well-controlled approximations based on molecular similarity

Matteus Tanha[1], Shiva Kaul[2], Alex Cappiello[1], Geoffrey J. Gordon[2] and David J. Yaron[1]
[1]Department of Chemistry
[2]Machine Learning Department
Carnegie Mellon University
Pittsburgh PA 15213

A means to take advantage of molecular similarity to lower the computational cost of electronic structure theory is proposed, in which parameters are embedded into a low-cost, low-level (LL) *ab initio* theory and adjusted to obtain agreement with a higher level (HL) *ab initio* theory. This approach is explored by training such a model on data for ethane and testing the resulting model on methane, propane and butane. The electronic distribution of the molecules is varied by placing them in strong electrostatic environments consisting of random charges placed on the corners of a cube. The results find that parameters embedded in HF/STO-3G theory can be adjusted to obtain agreement, to within about 2 kcal/mol, with results of HF/6-31G theory. Obtaining this level of agreement requires the use of parameters that are functions of the bond lengths, atomic charges, and bond orders within the molecules. The argument is made that this approach provides a well-controlled means to take advantage of molecular similarity in quantum chemistry.

## A. Introduction

*Ab initio* electronic structure theory has made great strides over the past few decades in developing methods with systematically improvable accuracy. For many classes of molecules, chemically accurate predictions can now be obtained[1,2]. A remaining challenge is lowering the computational cost such that accurate predictions may be obtained on large systems, such as those of interest in biological and materials applications. Ideally, the approximations invoked to lower the computation cost will have (*i*) a predictable level of accuracy and (*ii*) a means for systematic improvements in accuracy. An additional criterion that seems desirable is that the approximations have (*iii*) a form that can be justified in terms of an underlying principle. We will use these as the criteria for a "well-controlled" approximation. Molecular systems have two properties that provide the basis for approximations that lower computational costs: nearsightedness and molecular similarity. Nearsightedness implies that interactions become simpler at long-range[3], and can be replaced by increasingly coarse-grained multi-polar interactions. Algorithms that take full advantage of nearsightedness can achieve a computational cost that scale linearly with system size in the limit of large systems. Such linear-scaling algorithms have been developed for many *ab initio* methods[4–9]. The approximations introduced

44

in such algorithms are well controlled. There are typically parameters that define the degree to which interactions are simplified at long range, and these parameters can be adjusted to both estimate (criterion *i*) and reduce (criterion *ii*) the error introduced by the approximation. The approximations invoked in linear scaling, such as replacing charge distributions with multipoles at long range, also have a form that can be well justified in terms of the nature of Coulomb interactions (criterion *iii*). The other main aspect of molecular systems that can lead to a substantial reduction in computational cost is molecular similarity, whereby molecular fragments such as functional groups behave similarly in similar environments. Molecular similarity is the basis for molecular mechanics[10] and semiempirical quantum chemistry[11,12]. While such methods have had great success, they do not provide a well-controlled approximation scheme in the sense described above. The methods typically invoke specific forms for energy functions or effective Hamiltonians, without a scheme for systematically estimating (criterion *i*) or reducing (criterion *ii*) the error. The form of the approximation, such as the nature of the effective Hamiltonian, can also be difficult to justify in terms of underlying physical principles. Our goal is to develop well-controlled approximations that take advantage of molecular similarity.

To explore such approximations, we construct the problem in a manner that is amenable to machine learning techniques[13]. We embed parameters into a low level (LL) *ab initio* theory and adjust these to obtain agreement with results of a high level (HL) *ab initio* theory. We will refer to such models as machine learning-LL (ML-LL) models, to reflect that the LL model is being used as a functional form to be trained through machine learning techniques. The embedded parameters are associated with molecular fragments, and agreement between ML-LL and HL models is sought on data that spans a range of chemical systems over which molecular similarity can be plausibly assumed to hold. This approach has the potential to satisfy the above criteria for a well-controlled approximation. To achieve a predictable level of accuracy, we divide the data used to parameterize the model into a training and a test set. The model parameters are adjusted to obtain agreement on the training set, while monitoring the performance on the test set. The accuracy achieved on the test set thereby provides an estimate of the accuracy of the ML-LL model on systems that are sufficiently similar to those used to train the model (criterion *i*). The model can be systematically improved by increasing either the sophistication of the LL *ab initio* theory, by increasing the size of the basis or level of correlation, or by improving the means through which the parameters are embedded to form the ML-LL model, as is done below through the use of context sensitive parameters (criterion *ii*) .

Satisfying criterion *iii* is more challenging and likely requires advances in our understanding of the nature of molecular similarity. Here, we use minimal-basis Hartree-Fock theory (HF/STO-3G) as the basis for the ML-LL model and a spit-valence Hartree-Fock theory (HF/6-31G) as the HL theory. The HL theory differs from the LL by allowing the charge distribution to expand and contract in response to changes in the molecular geometry and the electrostatic environment. Our hypothesis is that the effects of this expansion and contraction behave similarly within a restricted class of molecules. If this holds, then we should be able to train a ML-LL model on a subset of molecules within that class and create a model that applies across the entire class. In addition, we embed the parameters in a manner that builds on intuitions regarding the effect the ML-LL model is attempting to capture. For instance, we invoke the intuition that the expansion and contraction is influenced by the charge on the atom. The training of the ML-LL model on HL data, as opposed to experimental data, is an advantage for satisfying criterion *iii,* since the nature of the differences between the LL and HL model are much better understood than the differences between the LL model and experiment.

Our approach is similar to traditional semiempirical quantum chemistry (SEQC) [11,12] in that parameters are embedded into a quantum mechanical Hamiltonian and adjusted to obtain agreement with available data. There are however a number of important distinctions. First, is the use of a training and test set to obtain an estimate of the error introduced by the approximation. A second distinction relates to the nature of the data used for the parameterization. Here, the model is parameterized to a large set of data generated from HL *ab initio* theory, including the expectation values of all operators that appear in the electronic Hamiltonian. Traditional SEQC attempts to obtain agreement with either a small set of experimental data, or with sparse data (geometries, heats of formation) obtained from HL *ab initio* theory. A third distinction is the meaning attributed to the embedded parameters. The parameters of SEQC are typically assumed to have some meaning, such as atomic ionization potentials or screened Coulombic interactions. Here, the parameters are viewed in a manner similar to the parameters embedded into a neural net, with little meaning being attached to the parameters themselves. The ML-LL model serves as a convenient functional form for an approximating function that is assumed to hold over some limited range. Finally, the model is assumed to be valid only for molecules that are sufficiently similar to those included in the training of the model. The intent is that models will be developed for different classes of molecules, making the approach more analogous to molecular mechanics, where parameters are often specific to particular fragments such as amino acids or polymer subunits.

The SCC-DFTB method[14,15] has some similarities with the current approach, including the use of detailed data from a higher-level theory to extract parameters for a LL model and the use of parameters tuned to specific classes of systems. However, our approach differs in the use of a training and test set to estimate error and in the general approach of using an ML-LL model as an approximating function that is trained to data with little regard for the meaning of the embedded parameters.

The choice of LL and HL theory for the current work is partly motivated by our previous work on empirical models. Features extracted from the results of a LL calculation were used to predict the results of a HL calculation. For instance, such a model was successful at predicting the two-electron density matrix, and thus the correlation energy, from the one-electron density matrix obtained from Hartree-Fock theory[16]. Another study considered the collinear reaction $H_2 + F \rightarrow HF + H$ in environments that strongly perturb the reaction energy profile[17]. A linear regression obtained chemical accuracy (<0.6 kcal/mol) in predicting the results of a HL calculation (QCISD/6-31G++**) using only inputs (energy and distributed multipoles) generated from a LL calculation (HF/3-21G). The error in such models was dominated by the extrapolation from small to large basis sets (3-21G $\rightarrow$ 6-31G++**), with much less error resulting from extrapolating to correlated theories (HF $\rightarrow$ QCISD). These results suggest that a key challenge in the development of empirical models is the extrapolation across basis set. This past work was empirical, in the sense that simple linear models were used to predict HL results from LL inputs. The current work is semi-empirical, in the sense that a modified Hamiltonian is used as the functional form in which to embed parameters. This model form may lead to substantially improved performance, especially given the success of traditional SEQC, which used only a handful of embedded parameters fit to a handful of experimental data. Our choice of LL and HL model is meant to explore the degree to which a semiempirical model can address the aspect that was most difficult in our past empirical models, the extrapolation across basis set.

The chemical systems considered here are saturated hydrocarbons. Section B-1 describes the data used to train and test the model. The form of the model is then described, including the embedding of parameters into the LL Hamiltonian (Section B-2), and addition of context sensitivity to these parameters (Section B-3). The training and performance of the model are then described in Section C.

## B. Methods

### B-1. Chemical Data

The data consists of the electronic structure of a hydrocarbon with varying geometries in a range of environments. Both LL (HF/STO-3G) and HL (HF/6-31G) data is generated. Random geometries are obtained by perturbing the equilibrium geometry with random distortions. These distortions are generated using a z-matrix that defines the molecular geometry in terms of bond lengths, angles, and dihedrals. A random number is then added to each value in the z-matrix, using a uniform random distribution with a width of $\pm 0.15$ Å for bond lengths, $\pm 6^o$ for bond angles, and $\pm 7^o$ for dihedral angles. Bond angles that are not explicitly included in the z-matrix thus have a range that is up to twice that of the angles included in the z-matrix. For ethane, the random variable is chosen to span the full range for the internal rotation angle.

The environment perturbs the electronic structure of the fragment in a manner that explores the types of perturbations that will be present in large molecules. This includes perturbations from external electrostatic potentials, due to other portions of the molecule or from solvent, and inductive effects from acceptors and donors. The environments consist of a cube, each corner of which holds a point charge. The length of each side of the cube is 12Å for methane and ethane and 14Å for propane and butane, with the molecule placed at its center. The magnitudes of the point charges are randomly generated using a uniform distribution between -25 and 25 amu, chosen to induce variations in the Mulliken charges on the C and H in methane and ethane that are similar to the charges induced on the methyl group in $CF_3CH_3$ (~0.2 amu).

For each pairing of a molecular configuration with an environment, we generate expectation values of each operator that appears in the Hamiltonian (total kinetic energy, interaction of the electron density with each nucleus, and total two-electron repulsion energy). Each data set consists of 10 configurations in 10 environments, corresponding to 100 calculations. Ethane has eight nuclei which, along with kinetic energy and two-electron energy, leads to 1000 data points. The goal of the model fitting is to get the 1000 values generated from the ML-LL model to agree with those from the HL theory.

### B-2. Effective Hamiltonian

Rather than use the *ad hoc* functional forms of traditional SEQC, we embed parameters into a LL (STO-3G) *ab initio* model. A systematic comparison of different embedding schemes will be presented in a later paper. Here, we choose a scheme that performs well and focus on issues related to the training and testing of the model.

For one-electron operators, we embed parameters by multiplying the matrix elements, $(i|h_1|j)$, by a multiplicative constant, (1+x), with x constrained to be greater than -1. For diagonal blocks, where i and j are atomic orbitals on the same atom, a different parameter is used for each atom type (C, H) and for each shell (1s, 2s, and 2p). For off-diagonal blocks, where i and j are on different atoms, modifications are included only between bonded atoms. For each bond, the atomic orbitals of C are transformed to create an $sp^3$ hybrid orbital directed along the bond, and the matrix element between the orbitals participating in this bond (1s for H, $sp^3$ for C) are multiplied by (1+x). (Note that only singly-bonded molecules are included here.) The parameter, x, used between bonded atoms is a function of the atom types participating in the bond. Integrals that are not modified according to the above rules retain their LL values, as opposed to being set to zero.

For the two electron operators, we use a multiplicative constant, (1+x), to modify the following classes of two electron integrals:
  *Diagonal:*      (ij|kl)  with i j k and l all on the same atom
  *Off-diagonal:*  (ij|kl)  with i j on one atom and k l on another atom
The diagonal element in which all orbitals are 1s has a single parameter for each atom type. For the (2s, 2p) shells of carbon, we use the form introduced by Slater and utilized in INDO theory[18] to express the on-atom integrals in terms of three parameters, $F^0$, $G^1$ and $F^2$. For the off-diagonal integrals, the values from the LL STO-3G theory are multiplied by a constant that depends on the two atom types. Between C and H, and between C and C, the integrals are modified only if there is a bond between the atoms. For integrals between H and H, the integrals are always modified. This approach to the two-electron operator has some similarities with the Zero Differential Overlap (ZDO) approximation, since the two-electron integrals being modified are those that are included in ZDO theories. However, unlike ZDO, the remaining two-electron integrals are retained, and set to the value they have in the unmodified LL theory.

To further illustrate the embedding scheme, Table 1 lists the parameters used for ethane.

**Table 1 List of parameter types for ethane.**

| Matrix element type | Atom types | Number of Parameters |
|---|---|---|
| Kinetic energy (KE) Diagonal | C | 3 (1s, 2s, 2p) |
| | H | 1 (1s) |
| Off-diagonal | C-H | 1 ($sp^3 - 1s$) |
| | C-C | 1 ($sp^3 - sp^3$) |
| Elec-nuclear interaction Diagonal | C | 3 (1s, 2s, 2p) |
| | H | 1 (1s) |
| Off-diagonal | C-H | 1 ($sp^3 - 1s$) |
| | C-C | 1 ($sp^3 - sp^3$) |
| Two-electron On-atom | C | 4 (1s, F0, G0, G2) |
| | H | 1 (1s) |
| Between atoms | C-H, C-C, H-H | 3 |
| TOTAL | | 20 |

### B-3.    Context sensitive parameters

We expect that a given set of parameters will be valid only over some limited range of molecules. We can extend this range by making the parameters functions of the current context of the molecule, where the context is extracted from the geometry and electronic density matrix. The current work considers the *ad hoc* context variables shown in Table 2. (The use of feature extraction methods to develop machine-learning derived contexts will be presented in a future work.) Below, these contexts are added sequentially. The first context is bond length, so we can determine the improvement obtained from including only geometry dependence in the model. The second context is the aspect of the electron density that seems most likely to be of relevance, with regards to the expansion and contraction of charge density present only in the HL model. For diagonal blocks, as charge is pushed onto (or pulled from) an atom, we expect the charge density to expand (or contract). For the off-diagonal blocks, the bond order may play a similar role of influencing the expansion and contraction of the electron density. The third context crosses these, using bond order for diagonal blocks and atomic charges for off-diagonal blocks.

**Table 2 *Ad hoc* context variables**

| Contexts for diagonal blocks |
|---|
| 1)    r: Average bond length to bonded atoms |
| 2)    q: Mulliken charge on the atom |
| 3)    bo: Average bond order to bonded atoms |
| **Contexts for off-diagonal blocks** |
| 1)    r: Bond length |
| 2)    bo: Bond order |
| 3)    q: Difference in charges on bonded atoms |

Since the density matrix is updated on each iteration of the Hartree-Fock (HF) solution process, the context can also be updated on each iteration and so integrate smoothly into the HF algorithm. However, for the fits shown here, the context variables are derived from HF/STO-3G and are not updated as the model is trained. In addition, the charges are those induced by the environment, as opposed to absolute charges.

Each of the parameters embedded in the Hamiltonian (Table 1) is made a linear function of the three context variables listed in Table 2. An exception is the parameter that modifies the two-electron integrals between hydrogen atoms: it becomes a linear function of only the bond-order. The restriction to bond order is motivated by the fact that these interactions are present between non-bonded atoms. The inclusion of bond order allows the model to make a small distinction between adjacent versus distant hydrogens. The resulting model has 78 embedded parameters.

## C. Results

As discussed in Section B-1, each data set consists of 10 molecular configurations coupled with 10 electrostatic environments, for a total of 100 molecule/environment pairs. The model is trained on such a data set for ethane. During training, performance is monitored on a test set, also for ethane, which has the same size as the training set. The model parameters are optimized using the trust-region reflexive algorithm[19,20], as implemented in MATLAB 2012a [21], with the objective being the RMS error of the training set. The error is computed for each operator, and each molecular environment pair,

$$Err_{mol,env}^{\hat{O}_i} = \left\langle \hat{O}_i \right\rangle_{mol,env}^{HL} - \left\langle \hat{O}_i \right\rangle_{mol,env}^{ML-LL}, \quad (1)$$

where $i$ ranges over kinetic energy (KE), the electron-nuclear interaction for each atom in the molecule, and the two-electron energy ($E_2$). We also consider the operator that sums all of these terms to give $E_{tot}$. Note that the energy of interaction with the environment is not included in $E_{tot}$. The environment is used to perturb the electronic distribution, and the model is adjusted to the self-energy of the molecule in the presence of such perturbations. The RMS error sums this over all environment/molecule pairs and over each operator,

$$RMS\ error = \sqrt{\frac{1}{N} \sum_{i,mol,env} \left( Err_{mol,env}^{\hat{O}_i} \right)^2}, \quad (2)$$

where $N$ is the total number of terms in the summation. In the fits shown below, the summation of Eq. (2) includes $E_{tot}$, with an optional weighting factor, $w$.

The errors of Eqs. (1) and (2) reflect disagreement between the ML-LL and HL models with regards to total energy. However, the absolute energy from a quantum chemical model has little meaning since it is only energy differences that can be measured experimentally. To capture the error associated with energy differences, we subtract the mean of the error,

$$Err'^{\hat{O}_i}_{mol,env} = Err_{mol,env}^{\hat{O}_i} - \overline{Err_{mol,env}^{\hat{O}_i}}, \quad (3)$$

where the mean is taken over all molecule/environment pairs, for each operator type. The RMS error is then,

$$RMS\ error' = \sqrt{\frac{1}{N} \sum_{i,mol,env} \left( Err'^{\hat{O}_i}_{mol,env} \right)^2}. \quad (4)$$

The error of Eq. (4) thus reflects the disagreement between the ML-LL model and HL model regarding changes in the operator expectation values arising from changes in either the geometry or environment of the molecule. Since there are sufficient parameters in the ML-LL model for the fitting algorithm to adjust the mean of the predictions to that of the HL model, Eqs. (2) and (4) are expected to be the same for the training set. However, Eq. (4) provides a more relevant measure of the performance on test sets.

Figure 1 shows the error as a function of iteration number for a weighting factor of $w=1$ on $E_{tot}$. The model is first optimized without the inclusion of context dependence in the parameters. Once convergence is achieved, the first level of context dependence is added to the model (contexts labeled 1 in Table 2), and so on. The fit shown here used tight convergence criteria ($10^{-10}$ relative change in either the parameters or the RMS error, or 200 maximum iterations at each stage). The objective tends to decrease smoothly, such that in later fits, we stop the optimization when there are 4 successive steps in which the performance on the test set drops by less than 0.01 kcal/mol. In most cases, this stopping criteria is met when performance on the test set degrades for four successive steps. This criteria would stop the optimization before the drop in RMS error that occurs near iteration 325 in Figure 1. This may indicate the need for a global optimization procedure[22]. However, this drop does not lead to improved performance on the total energy and so does not suggest a need to alter our stopping criteria. The performance on the test set roughly tracks that of the training set, with a final performance for the total energy only 30% higher than the training set, 4 kcal/mol average for the training set as opposed to 3 kcal/mol for the test set.

Figure 2 shows the average error for each operator type. The error for these operators is substantially higher than that of the total energy in Figure 1, indicating that there is considerable cancellation of error when the operators are summed to give the total energy. The cancellation of errors is likely related to the variational nature of the Hartree-Fock method, which minimizes the total energy under the constraint of a single Slater determinantal wavefunction. Since only the total energy is minimized, the better performance observed for $E_{tot}$ versus other operators may not be surprising.

The performance of the model of Figure 1 on molecule types not included in the training is shown in Figure 3 through Figure 5. Figure 3 shows the error in the absolute energy, Eq. (2), in which case the performance is quite poor for molecules not included in the fit. However, the absolute energy of a quantum chemical model has little meaning since it is only energy differences that can be measured experimentally. Figure 4, shows the error of Eq. (4), which reflects the performance expected for such energy differences. The performance on molecules not included in the fit is comparable to that obtained on the test set of ethane molecules. This verifies that a model trained on one type of molecule can be transferred to other similar molecules. Figure 5 shows the error of Eq. (4), summed over all operators except that correspond to $E_{tot}$. The results show that performance on molecules types not included in the training is comparable to the performance of the ethane test set.

The better performance obtained for $E_{tot}$ than individual operators indicates a substantial cancellation of errors upon addition of operators to obtain the total energy. This suggests that improved performance for $E_{tot}$ may be obtained by weighting $E_{tot}$ more strongly in the objective function of Eq. (2). The performance on energy differences, computed via Eq. (4), is shown as a function of such a weighting parameter in Figure 6 and Figure 7. Figure 6 shows that the performance for the total energy tends to improve with increased weighting of the total energy. This includes performance on molecule types not included in the training of the model, although the improvement on the training set is substantially larger than that on the test sets. Figure 7 shows that the improved performance on the total energy comes at a cost to performance on the individual operators, with the error becoming 1000's of kcal/mol when individual operators are not included in the fit (a weight of infinity). Fits to just the total energy therefore obtain good performance on the total energy, but very poor performance for individual operators. This poor performance on individual operators does not appear to harm transfer of model parameters between molecular systems, as reasonable performance is obtained for the total energy of molecule types not included in the training of the model, even when the molecule is trained on just total energy (see results for a weight of infinity in Figure 6).

**Table 3 Detailed errors (in kcal/mol) from the fit in Figure 6 corresponding to w=10.**

|  | Error in Energy Eq. (2) | | | | | Error in Energy Differences Eq. (4) | | | | |
|---|---|---|---|---|---|---|---|---|---|---|
|  | KE | $EN_H$ | $EN_C$ | $E_2$ | $E_{tot}$ | KE | $EN_H$ | $EN_C$ | $E_2$ | $E_{tot}$ |
| **Ethane(train)** | | | | | | | | | | |
| Initial | 938 | 45 | 716 | 311 | 545 | 126 | 21 | 86 | 112 | 17 |
| No Context | 25 | 27 | 25 | 29 | 7.6 | 24 | 26 | 25 | 28 | 7.6 |
| Context | 6.6 | 3.6 | 5.4 | 7.1 | 0.7 | 6.6 | 3.6 | 5.3 | 7 | 0.7 |
| **Ethane(test)** | | | | | | | | | | |
| Initial | 904 | 51 | 693 | 352 | 542 | 142 | 23 | 93 | 111 | 15 |
| No Context | 33 | 28 | 31 | 42 | 7.2 | 32 | 28 | 31 | 42 | 7.2 |
| Context | 11 | 4.2 | 9.2 | 11 | 2.1 | 11 | 4.2 | 9.2 | 11 | 2.1 |
| **Methane** | | | | | | | | | | |
| Initial | 424 | 47 | 717 | 161 | 279 | 63 | 22 | 67 | 51 | 8.8 |
| No Context | 197 | 28 | 482 | 213 | 60 | 13 | 28 | 44 | 18 | 6.2 |
| Context | 38 | 5.2 | 144 | 110 | 2.2 | 8 | 5.1 | 13 | 8.5 | 2.2 |
| **Propane** | | | | | | | | | | |
| Initial | 1456 | 39 | 737 | 372 | 831 | 81 | 17 | 56 | 66 | 6.1 |
| No Context | 182 | 21 | 252 | 248 | 55 | 18 | 20 | 206 | 51 | 5.2 |
| Context | 49 | 4.3 | 97 | 250 | 5.9 | 10 | 4.3 | 12 | 14 | 1.3 |
| **nButane** | | | | | | | | | | |
| Initial | 1959 | 40 | 734 | 503 | 1104 | 82 | 18 | 65 | 80 | 7.7 |
| No Context | 379 | 22 | 328 | 608 | 101 | 19 | 22 | 200 | 64 | 4.7 |
| Context | 136 | 6.3 | 196 | 657 | 19 | 14 | 6.3 | 27 | 32 | 2.2 |
| **tButane** | | | | | | | | | | |
| Initial | 1893 | 45 | 718 | 554 | 1109 | 119 | 18 | 66 | 90 | 8 |
| No Context | 327 | 22 | 416 | 535 | 91 | 34 | 22 | 343 | 47 | 4.9 |
| Context | 107 | 5.7 | 195 | 680 | 19 | 16 | 5.7 | 23 | 37 | 1.7 |

A weighting of 10 in Figure 6 and Figure 7 reduces the error in the total energy on the test molecules, while having little impact on the errors for the individual operators. Table 3 summarizes the results obtained from this fit. The left portion of the table shows errors in the absolute energies, while the right shows errors in energy differences. The rows labeled *initial* refer to differences between STO-3G (LL) and 6-31G (HL). The *initial* disagreement regarding energy differences, Eq. (4), is large for individual operators and considerably smaller for the total energy, $E_{tot}$. This indicates substantial cancellation of errors between the operators of STO-3G and 6-31G. The use of constant parameters, labeled *no context* in the table, reduces the error in the total energy by about a factor of two for both the train and test data sets, while substantially improving the agreement on individual operators. The decrease of only a factor of two in the error associated with the total energy speaks to the complex nature of the data set being explored here. The environments are perturbing the electron density in ways that cannot be captured by simply scaling the matrix elements of the minimal basis Hamiltonian. When the model parameters are made functions of context, the model improves substantially. With context included, the error for $E_{tot}$ of test molecules is reduced by a factor of between 3.5 and 7 relative to the initial error. Performance on individual operators is also improved substantially, although these remain considerably larger than the error in $E_{tot}$.

## D. Conclusions

This work explores a form of semiempirical model in which parameters are embedded into a LL *ab initio* theory and adjusted to obtain agreement with a HL *ab initio* theory. The results suggest that the approach meets the criteria discussed in Section A for providing a well-controlled approximation based on molecular similarity.

The first criterion is that the approximation introduces a predictable level of accuracy. Estimates of the error introduced by the approximation are obtained from the performance of the model on test data. Here, we trained a model to data on ethane and monitored convergence of the training process on a test set containing also data on just ethane. The performance on this ethane test set correlated well with the performance seen on molecules (methane, propane, and butane) not included in the training process. This suggests that the performance on test data provides a reliable measure of the error introduced by the approximation.

The second criterion is that the model has a means for systematically improving the accuracy. The general approach explored here may be improved by increasing either the sophistication of the LL *ab initio* theory into which the parameters are embedded, or by improving the means through which the parameters are embedded. The current implementation considered only the latter, by making the parameters functions of the molecular context such as bond lengths, atomic charges and bond orders. Incremental addition of the context variables to the model lead to steady improvements in the accuracy of the model, as judged by performance on test data.

The third criterion is that the approximations have a form that can be justified in terms of an underlying principle. It is less clear that the approach implemented here meets this criterion. The use of charge and bond-order as context variables is connected to the intuition that the charge density should expand or contract in response to changes in electron density on an atom or within a bond. The success of these context variables in improving the fit of the ML-LL model supports this intuition.

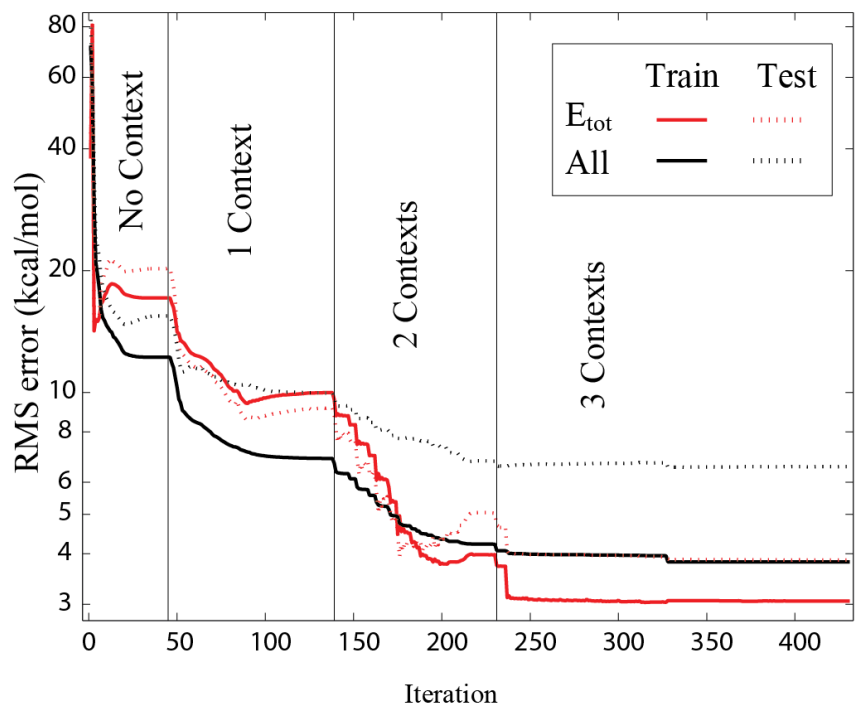

**Figure 1** Training of the ML-LL model on an ethane data set (solid lines), while monitoring performance on an ethane test set (dotted lines). The objective is the RMS error, Eq. (2), summed over all operators including $E_{tot}$ (black lines). The RMS error, Eq. (2), for just the total energy, $E_{tot}$, (red lines) is also shown. The vertical lines show addition of context variables to the model (Table 2).

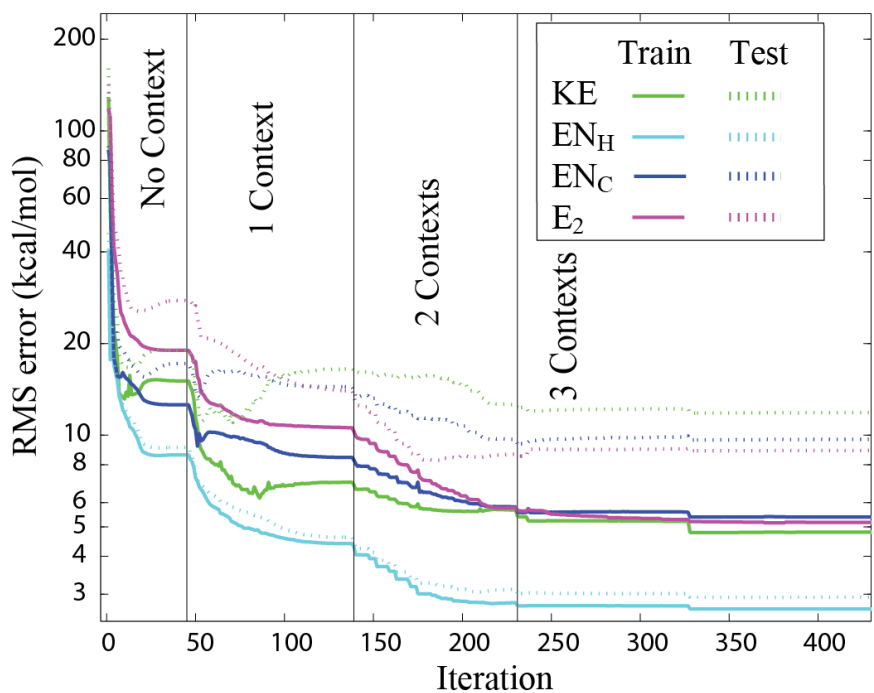

**Figure 2** Error in the individual operators, excluding $E_{tot}$, for the optimization of Figure 1. The RMS error is obtained using Eq. (2), with summations limited to a particular operator type.

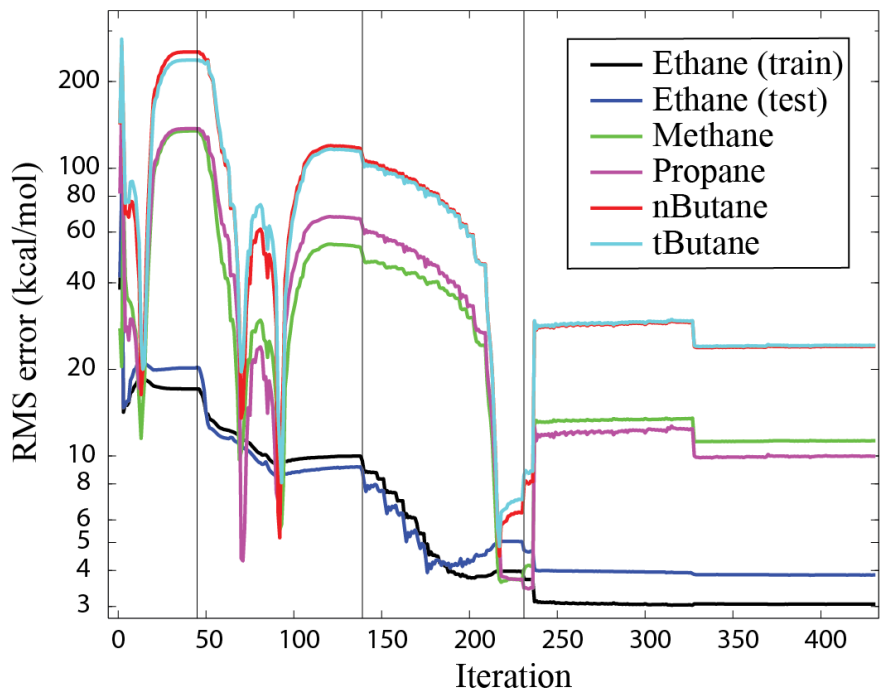

**Figure 3** RMS error, Eq. (2), in the total energy, $E_{tot}$, for a variety of molecules, using the ML-LL model of Figure 1.

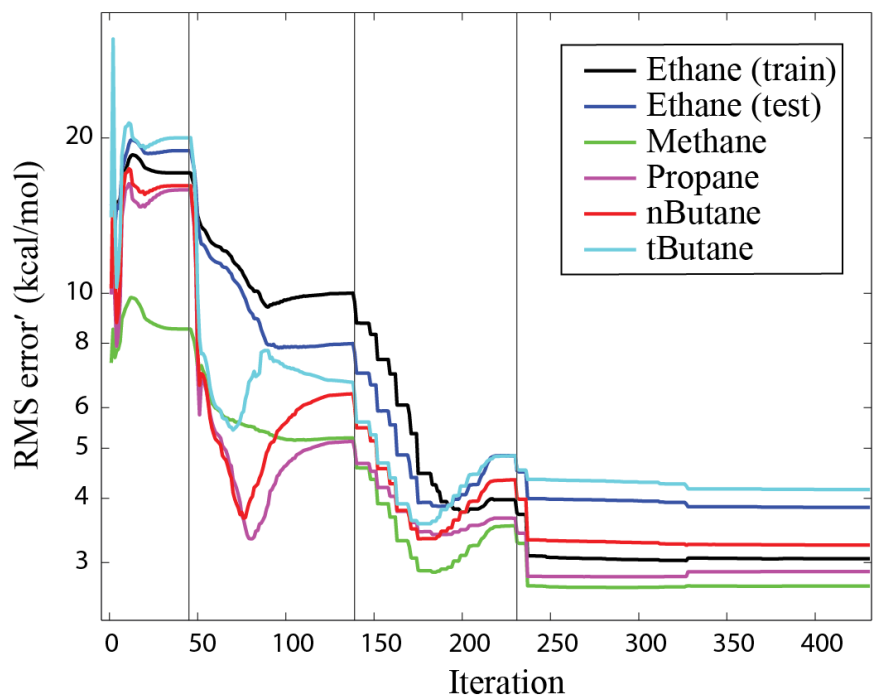

**Figure 4** RMS error′, Eq. (4), in the total energy, $E_{tot}$, for a variety of molecules, using the ML-LL model of Figure 1. This error reflects performance on the change in total energy arising from changes in geometry or environment.

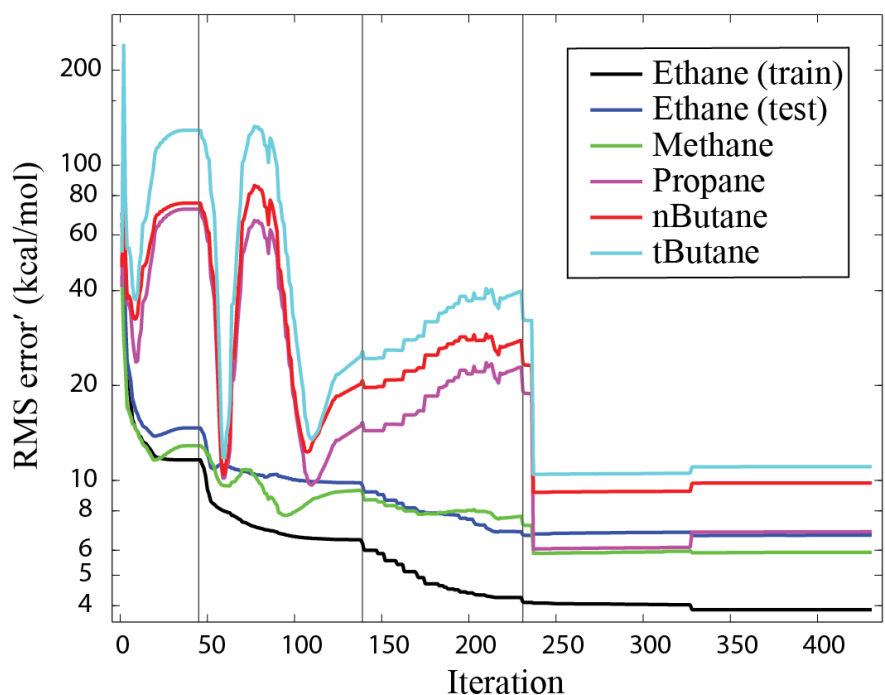

**Figure 5** RMS error′ associated with individual operators for the fits of Figure 4, computed using Eq. (4) with summation over all operators except $E_{tot}$.

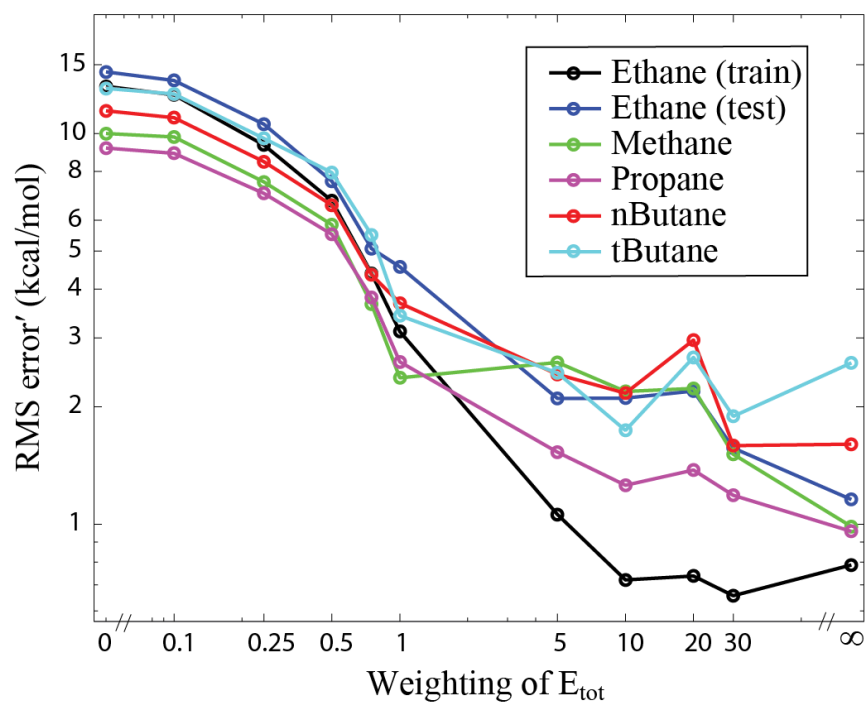

**Figure 6** RMS error′, Eq. (4), in total energy of various molecules, obtained from fitting the ML-LL model to ethane, with the total energy multiplied by a weighted factor.

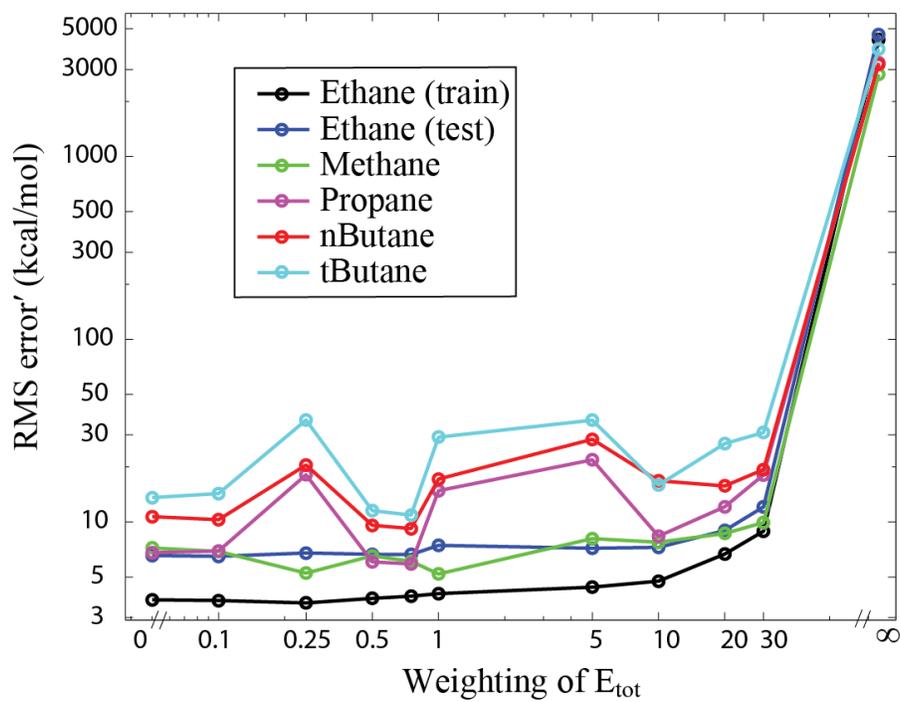

**Figure 7** RMS error′ associated with individual operators for the fits of Figure 6, computed using Eq. (4) with summation over all operators except $E_{tot}$.